\documentclass[conference]{IEEEtran}
\usepackage{cite}
\usepackage{amsmath,amssymb,amsfonts}
\usepackage{graphicx}
\usepackage{textcomp}
\usepackage{xcolor}
\usepackage{acronym}
\usepackage{changepage}
\usepackage{algorithm}
\usepackage[noend]{algpseudocode}
\usepackage{balance}

\algblock{Input}{EndInput}
\algnotext{EndInput}
\algblock{Output}{EndOutput}
\algnotext{EndOutput}
\newcommand{\Desc}[2]{\State \makebox[8em][l]{#1}#2}
\newcommand{\pH}{\text{pH}}
\newcommand{\pKa}{\text{p}K_a}

\begin{document}

\title{A clustering-based biased Monte Carlo approach to protein titration curve prediction}

 \author{\IEEEauthorblockN{Arun V. Sathanur}
 \IEEEauthorblockA{Pacific Northwest National Laboratory \\ 
 Seattle, WA, USA \\ 
 arun.sathanur@pnnl.gov}
 \and
 \IEEEauthorblockN{Nathan A. Baker}
 \IEEEauthorblockA{Pacific Northwest National Laboratory \\ 
 Richland, WA, USA \\ 
 nathan.baker@pnnl.gov}
 }

\maketitle

\begin{abstract}
In this work, we developed an efficient approach to compute ensemble averages in systems with pairwise-additive energetic interactions between the entities.
Methods involving full enumeration of the configuration space result in exponential complexity.
Sampling methods such as Markov Chain Monte Carlo (MCMC) algorithms have been proposed to tackle the exponential complexity of these problems; however, in certain scenarios where significant energetic coupling exists between the entities, the efficiency of the such algorithms can be diminished.
We used a strategy to improve the efficiency of MCMC by taking advantage of the cluster structure in the interaction energy matrix to bias the sampling.
We pursued two different schemes for the biased MCMC runs and show that they are valid MCMC schemes.
We used both synthesized and real-world systems to show the improved performance of our biased MCMC methods when compared to the regular MCMC method. 
In particular, we applied these algorithms to the problem of estimating protonation ensemble averages and titration curves of residues in a protein.
\end{abstract}


\section{Introduction}
Machine learning (ML) is playing an increasingly larger role in helping shape our understanding of complex biological phenomena through a number of related disciplines such as biophysics, biochemistry, molecular dynamics and bioinformatics to name a few~\cite{ramanathan2015large,botu2015adaptive,baldi2001bioinformatics}.
One recurring problem in many of these disciplines is the assignment of states to pairwise-interacting entities so as to minimize the total energy of the system and to compute the ensemble averages of properties that govern the behavior of the system.
The brute-force complexity of such problems is $k^N$ where $k$ is the number of possible states per entity and $N$ is the total number of entities.
The interactions of these $N$ entities can be described by a connected graph and the discrete optimization moves cannot be performed in isolation.
One such problem involves titration state assignment to the amino acid residues in a protein.

The protein titration problem involves prediction of the protonation ensemble averages for each residue as a function of $\pH$; residue $\pKa$ values can be derived from the resulting function~\cite{Purvine2016}.
When residues are found in a protein, their $\pKa$ values can vary significantly from isolated ``ideal'' residues due to the influence of the interacting protein residues.
The energy function $G(P)$ in equation \ref{equn:total_energy} is represented as the sum of the intrinsic energies of the titratable residues and the pair-wise interaction terms between residues residing on a rigid backbone.
The interaction energy terms are dependent on the titration states where we assume that each titratable residue has one of two states:  protonated or de-protonated.
\begin{equation}
    \begin{split}
        G(P) = \sum_{i=1}^{i=N} \gamma_i ln(10)kT\left(\pH-\text{p}K^{int}_{a_i}\right) + \\
        \sum_{\substack{i=1 \\ i > j}}^{i=N}\sum_{j=1}^{j=N}\gamma_i \gamma_j U^{int}(P_i,P_j)
    \end{split}
    \label{equn:total_energy}
\end{equation}
Here $N$ is the total number of titratable residues.
$P_i$ denotes whether the residue \textit{`i'} is protonated or deprotonated, $U^{int}(P_i,P_j)$ denotes the interaction energy between residues $i$ and $j$ based on the states $P_i$ and $P_j$.
$P$ denotes the $N$ dimensional vector that describes the system configuration; the $i^{th}$ element of this vector denotes the state of the $i^{th}$ residue. The value $\text{p}K^{int}_{a_i}$ is the intrinsic $\pKa$ of the residue $i$ and $\gamma_i$ indicates whether the residue is charged ($\gamma_i = 1$) or uncharged ($\gamma_i = 0$).

The property is a function $f$ of the configuration vector $P$, the quantity of interest is the ensemble average
\begin{equation}
    \bar{f} = \frac{\sum_{m=1}^{2^N}f\left(P_m\right)\exp\left(-\beta G\left(P_m\right)\right)}{\sum_{m=1}^{2^N}\exp\left(-\beta G\left(P_m\right)\right)}
    \label{equn:mcmc_exp}
\end{equation}
Here $\beta$ is a constant given by the reciprocal of the product of the Boltzmann constant and the temperature in absolute scale.
The protein titration curve prediction is the value of $\bar{\gamma}_i$ as a function of $\pH$.
This work is concerned with an efficient and accurate estimation of the titration curves for all titratable residues in a protein where the interaction energy matrix $U^{int}$ has a good cluster structure. 

This work includes the following contributions:
\begin{itemize}
	\item We propose a biased MCMC scheme to efficiently estimate ensemble averages by leveraging the cluster structure of the interaction energy matrix.
	\item Through theoretical analyses, we justify our sampling schemes and show why the proposed schemes represent valid MCMC runs.
	\item Using experiments on synthetic and real-world systems, we show improvement over MCMC techniques that do not exploit the cluster structure.
\end{itemize}

Section \ref{sec:rel} describes the related work in estimating ensemble averages for the protein titration problem, Section \ref{sec:mcmc} describes the standard MCMC procedure and introduces our biased MCMC approach.
Section \ref{sec:theory} provides details on the algorithms and demonstrates that these are valid MCMC methods.
Section \ref{sec:syn} describes experiments on synthetic protein-residue networks and also explores the effect of clustering quality on the performance of the algorithm.
Section \ref{sec:real} then details the experiments on real-world datasets.
Section \ref{sec:conclu} concludes the paper with pointers on the future work.

\section{Related work}
\label{sec:rel}
Tanford and Roxby~\cite{tanford1972interpretation} developed a mean-field approach to compute the titration curves, approximating interactions by an effective average quantity to reduce computational cost.
The mean-field approach works well when the interaction energy terms are small compared to the intrinsic terms.
It therefore  performs poorly in the presence of strong energetic interactions between titration sites.
Bashford and Karplus developed an improved reduced-site approximation \cite{bashford1991multiple} by eliminating those residues that are almost-always protonated or deprotonated for a given $\pH$ and thereby improving the computational performance.
Related approaches use cluster structures within a protein-residue system and other approximations to the direct calculation of the partition function \cite{gilson1993multiple, myers2006simple}.
Alternatively, methods such as PROPKA simplify several steps such as the computation of the difference in solvation free-energy by using simplified energy calculations and heuristic methods for $\pKa$ estimates \cite{Sondergaard2011,  Olsson2011, olsson2011protein}.

MCMC approaches to titration state prediction naturally incorporate interactions between the residues through the total energy calculation. 
Slow MCMC convergence is driven by several factors, particularly the presence of energetically coupled clusters of titratable residues whose states change together \cite{beroza1991protonation}.
Some authors \cite{beroza1991protonation, rabenstein1998calculation} have addressed this convergence problem by allowing pairs and triplets of sites to simultaneously change their  states.
Our approach builds on this idea and extends the approach to arbitrary clusters as expressed by the data. 

Finally, the authors in a recent work \cite{Purvine2016} use a computer-vision inspired graph-cut optimization algorithm which finds lowest-energy titration states in a computationally efficient manner.
The authors use the optimum state to compute a mean-field titration curve, similar to methods used by Tanford and Roxby~\cite{tanford1972interpretation}, and subject to the limitations described above.
This approach requires a submodularity property of the energy interactions which may not be satisfied by all the residues.
Residues with energy interactions that are not submodular must have states assigned using brute-force or MCMC approaches.

\section{Ensemble average calculations and the MCMC algorithm}
\label{sec:mcmc} 

The protein titration problem described above closely resembles the popular Ising model that has been extensively studied \cite{chandler1987introduction}.
Ensemble average of system properties can be calculated by a traditional MCMC algorithm where the system starts with a random configuration and, at each move, one residue is selected at random.
The energy difference is calculated by flipping the protonation state of the selected residue.
The move is accepted with a probability equal to $min\left(1,e^{-\beta \Delta G}\right)$ where $\Delta G$ is the energy difference for the state change.
Thus, configurations leading to lower energy are always accepted while those resulting in increase in energy are accepted with a probability less than 1.
The Metropolis algorithm is first run for $T_1$ steps called burn-in allowing the system to equilibrate and, following that, sampling is done for $T_2$ steps to compute ensemble averages.

Previous authors \cite{beroza1991protonation} have noted that the traditional MCMC scheme of changing states one residue at a time, may--in certain cases--never sample states that require multiple transitions.
Therefore, they introduced moves that allowed changing two states at a time.
We generalized this idea to exploit the cluster structure present in protein residue interaction networks \cite{vijayabaskar2010interaction}.
While our focus was protein residue networks, the approach is general enough to be applied in any Ising-like problem where the interaction energy matrix has cluster structure.
The basic idea is to decompose the set of titratable residues into a set of clusters based on the interaction energies.
Residues within each of the clusters have a stronger coupling between each other than interactions between residues belonging to different clusters.
Allowing correlated state changes in residues that form strongly coupled clusters can avoid visiting energetically infeasible states and hence can accelerate the convergence of the MCMC runs.

Using the matrix of the pairwise interaction energies $U^{int}$, we first decomposed the set of $N$ titratable residues into $l$ (energetically strong) clusters $C_1,C_2,...,C_l$ with $n_1,n_2,...,n_l$ members each.
We require that $n_i \ge 2, 1 \le i \le l$, i.e., there are at-least two residues per cluster.
The remainder of the residues that are not strongly coupled with each other form a separate weak cluster $C_{l+1}$ with $n_{l+1} = N - \sum_{i=1}^{i=l}n_i$.
We used brute-force enumeration on each of these $l$ strong clusters, calculating the $k_r$ lowest-energy configurations for each cluster $C_r$ and storing them.
We described in a Section \ref{sec:theory} how the $C_{l+1}$ is treated differently from the energetically strong clusters $C_i$, $1 \le i \le l$.

\section{Details of the biased MCMC method}
\label{sec:theory} 

Our approach initially breaks the joint probability mass function (PMF) of the configuration space into $l+1$ blocks of correlated distributions with no correlations between the blocks.
This is similar to the clique decomposition of joint distributions in undirected graphical models \cite{koller2009probabilistic} with the cliques being replaced by non-overlapping clusters.
This decomposition allows us to cast configuration sampling in terms independent sub-sequences corresponding to the residues that constitute each of the clusters.
Since the individual cluster sizes are not too big (owing to the fact that strong interactions are confined to a small number of nearest neighbors in 3D space \cite{estrada2010universality}), the total number of configurations under consideration in each energetically strong cluster $C_r$ can be restricted to the most probable $k_r$ each (or lowest $k_r$ in terms of energy).
A brute-force strategy based on energy cut-off helps enumerate the sub-sequences of the configuration vectors corresponding to each of the clusters.
We sampled the $p^{n_{l+1}}$ configurations for the weakly-coupled cluster $C_{l+1}$, uniformly at random, because the number $n_{l+1}$ could potentially be too large for brute-force enumeration. 
The ratio of the configuration space sizes explored by the biased MCMC approach ($\left|\Omega_b\right|$) against the regular MCMC approach ($\left|\Omega_b\right|$) is given by
\begin{equation}
    \frac{\left|\Omega_b\right|}{\left|\Omega_r\right|} = \frac{\left(\prod_{r=1}^{l}k_r\right)p^{N - \sum_{r=1}^{i=l}n_r}}{p^N} = \prod_{r=1}^{l}\left(\frac{k_r}{p^{n_r}}\right).
    \label{equn:ratio}
\end{equation}
A valid state in the reduced configuration space consists of adjacent subsequences corresponding to the clusters $C_1$ through $C_{l+1}$.
For cluster $C_r$, where $r$ goes from $1$ through $l$, the subsequences will correspond to one of configurations from the top-$k_r$ considered for that cluster.
For the last cluster $C_{l+1}$, the corresponding subsequence will be one of the possible $p^{n_{l+1}}$ configurations sampled randomly as in traditional MCMC approaches.
Although we sample the subsequences from the individual clusters in an independent fashion, we account for the correlations across clusters by computing energies from Equation \ref{equn:total_energy} that include all the cross-cluster interactions needed for accurate energy computation.

\subsection{Satisfying ergodicity and detailed balance conditions}
For a MCMC algorithm to sample the configurations from the underlying joint distribution, two conditions need to be satisfied:
\begin{itemize}
\item Ergodicity: It should be possible to go from any configuration to any other configuration in a finite number of steps, and
\item Detailed Balance: The equilibrium rate of transitions into any configuration $P_{\mu}$ must be equal to the rate of transitions out of $P_{\mu}$.
\end{itemize}
Let $\pi\left(P_{\mu}\right)$ denote the steady-state (w.r.t.~to the Markov chain) probability of the configuration $P_{\mu}$.
$\pi\left(P_{\mu}\right)$ should be the Boltzmann probability under equilibrium conditions.
Let $T\left(P_{\mu},P_{\nu}\right)$ denote the proposed transition probability from $P_{\mu}$ to $P_{\nu}$ for the simulated Markov chain as part of the MCMC procedure.
Also let $A\left(P_{\mu},P_{\nu}\right)$ denote the acceptance probability of this proposed transition.
Then  the following equation needs to be satisfied for all pairs of states $P_{\mu} \text{ and } P_{\nu}$ to ensure the detailed balance:
\begin{multline}
    \pi\left(P_{\mu}\right)T\left(P_{\mu},P_{\nu}\right)A\left(P_{\mu},P_{\nu}\right) \\
    = \pi\left(P_{\nu}\right)T\left(P_{\nu},P_{\mu}\right)A\left(P_{\nu},P_{\mu}\right) 
    \label{equn:6}
\end{multline}
If both of the conditions in Equation \ref{equn:7} are satisfied, then the steady-state simulated Markov Chain samples the Boltzmann probability.
\begin{subequations}
    \label{equn:7} 
    \begin{eqnarray}
    T\left(P_{\mu},P_{\nu}\right) &=& T\left(P_{\nu},P_{\mu}\right) \label{equn:7a}  \\
    \frac{A\left(P_{\mu},P_{\nu}\right)}{A\left(P_{\nu},P_{\mu}\right)} &=& e^{-\beta\left(G(P_{\nu})-G(P_{\mu})\right)}    \label{equn:7b} 
    \end{eqnarray}
\end{subequations}

There are multiple ways to design the transitions in our approach; we considered two schemes.
Approach 1 first chooses a cluster uniformly at random and then chooses a sub-sequence configuration within that again uniformly at random.
Approach 2 chooses sub-sequence configurations uniformly at random in all the clusters. 

\subsection{Approach 1}

In this approach, the transition consists of two steps:
\begin{itemize}
    \item First choose a cluster $C_r$ at random. 
    \item If $r \in [1,l]$, then perturb the sub-sequence corresponding to $C_r$ by choosing a configuration uniformly at random from the $(k_r-1)$ configurations excluding the current sub-sequence.
    If $r = (l+1)$, then choose a residue at random in the cluster $C_{l+1}$ and flip its state.
\end{itemize}
Consider a valid initial state $P_{\mu}$ for a biased MCMC run.
By using one perturbation as described above, we end up in an overall system configuration denoted by $P_{\nu}$.
The transition probability for this can be easily seen as
\begin{equation}
    T\left(P_{\mu},P_{\nu}\right) = T\left(P_{\nu},P_{\mu}\right) =
    \begin{cases}
        \frac{1}{\left(l+1\right)\left(k_r-1\right)}, & \text{if $r \in [1.l]$}. \\
    	\frac{1}{\left(l+1\right)n_{l+1}}, & \text{$r = l+1$}.
 	\end{cases}
\label{equn:8}
\end{equation}
This demonstrates the symmetric nature of the transition proposal probability for the chain.
The Markov chain is ergodic because there is a finite probability of choosing any sub-sequence configurations in the clusters $1$ through $l$ and a finite probability of reaching any sub-sequence configuration in cluster $l+1$ in a finite number of steps.
Note that we exclude the current sub-sequence so that the probability of the transition $\left(P_{\mu}\rightarrow P_{\mu}\right)$ is $0$.
Thus Approach 1 represents a valid MCMC procedure.

\subsection{Approach 2}
In the second approach, an overall transition is obtained by applying transitions to all the $l+1$ subsequences corresponding to the clusters identified:
\begin{itemize}
    \item For each cluster $C_r$ where $r \in [1,l]$, perturb the sub-sequence corresponding to $C_r$ by choosing a configuration uniformly at random from the $k_r$ configurations
    \item For the cluster $C_{l+1}$, choose a residue at random in the cluster $C_{l+1}$ and flip its state.
\end{itemize}
Using similar arguments as in the case of Approach 1, we can show that Approach 2 also represents a valid MCMC procedure.

\subsection{A note on the clustering methods}
A weighted interaction graph is defined by the absolute value of the interaction energy matrix; this graph is partitioned into mutually exclusive clusters.
There are two major challenges for this step.
First, the cut edges edges need to be minimized: the entries that straddle neighboring clusters since minimizing the number of cut edges allows for a better sampling accuracy.
Second, the size of each partition should be balanced as much as possible in order to achieve computational efficiency.
To address both these issues, we used the hierarchical graph partitioning algorithm \emph{METIS} \cite{metis}.
The user must provide the number of partitions ($l$) as a hyper-parameter; the weak cluster $C_{l+1}$ is not used in this clustering strategy.
Other clustering strategies, such as the connected components method~\cite{vijayabaskar2010interaction}, can use the weak cluster $C_{l+1}$. 

\subsection{The full biased MCMC algorithm}
This section outlines the full algorithm to execute the biased MCMC approaches.
The version of the algorithm we present uses Approach 1 for the perturbation of the states within the MCMC procedure and the \emph{METIS} algorithm to determine the set of clusters.
We used an adaptive strategy to determine the values of $k_r$ per cluster based on a cut-off value such as retaining the energy states up to $5k_{B}T$ from the energy of the lowest state in that cluster.
\begin{algorithm}
    \small
    \caption{Biased MCMC}
    \begin{algorithmic}[1]
    \Input
        \Desc{$pHVals$}{List of pH values}
        \Desc{$pK_a^{int}$, $U^{int}$}{Self and interaction energies}
        \Desc{k}{No. clusters desired (Optional)}
        \Desc{$k_r$}{No. lowest energy states in $C_r$ (Optional)}
        \Desc{$N_1$,$N_2$}{Burn-in / Sampling steps}
    \EndInput
    \Output
        \Desc{$\bar{f_i}$ for $i$=$1$ to $N$ }{ Average protonation fractions}
    \EndOutput
    \State Construct a weighted, undirected graph $G$, with residues as nodes 
    \State Set weight $W_{ij}$ between nodes $i$ and $j$ as $\left|U^{int}(i,j)\right|$  
    \State Use a clustering algorithm to decompose $G$ into a set of $l$ clusters
    \For{$pH$ in $Pvals$}
        \State $S_r \Leftarrow$ lowest $k_r$ energy configurations in $C_r$ via enumeration
        \State $P \Leftarrow$ $\left[P_1 P_2 .... P_r,,,P_l\right]$ where $P_r$ is chosen uniformly random from $S_r$
        \For{$i$ = $1$ to $N$}
	        \State $\bar{f_i}(pH) = 0.0$
        \EndFor
        \For{step = 1 to $N_1+N_2$}
            \State Choose cluster $C_r$ uniformly from the set of $l$ clusters
            \State Sample $P^{'}_r$ uniformly at random from $S_r-{P_r}$
            \State $P^{'} \Leftarrow$ $\left[P_1 P_2 .... P^{'}_r,,,P_l\right]$ (only the state for $C_r$ is altered)
            \State $a$ = $min(1.0,\exp\left(-\left(\frac{\left(G(P^{'})-G(P)\right)}{k_BT}\right)\right))$
            \State $P = P^{'}$ with probability $a$
            \If{step $\ge$ $N_1$}
                \For{$i$ = $1$ to $N$}
                    \If{$P(i) = 1$}
		                \State $\bar{f_i}(pH) = \bar{f_i}(pH) + 1.0$
	                \EndIf
	            \EndFor
            \EndIf
        \EndFor
        \For{$i$ = $1$ to $N$}
	        \State $\bar{f_i}(pH) = \frac{\bar{f_i}(pH)}{N_2}$
        \EndFor
    \EndFor
    \end{algorithmic}
    \label{algo:biased_mcmc}
\end{algorithm}

\section{Experiments on a synthetic systems}
\label{sec:syn}

\begin{figure}[!htbp]
    \begin{centering}
        \includegraphics[width=0.9\columnwidth]{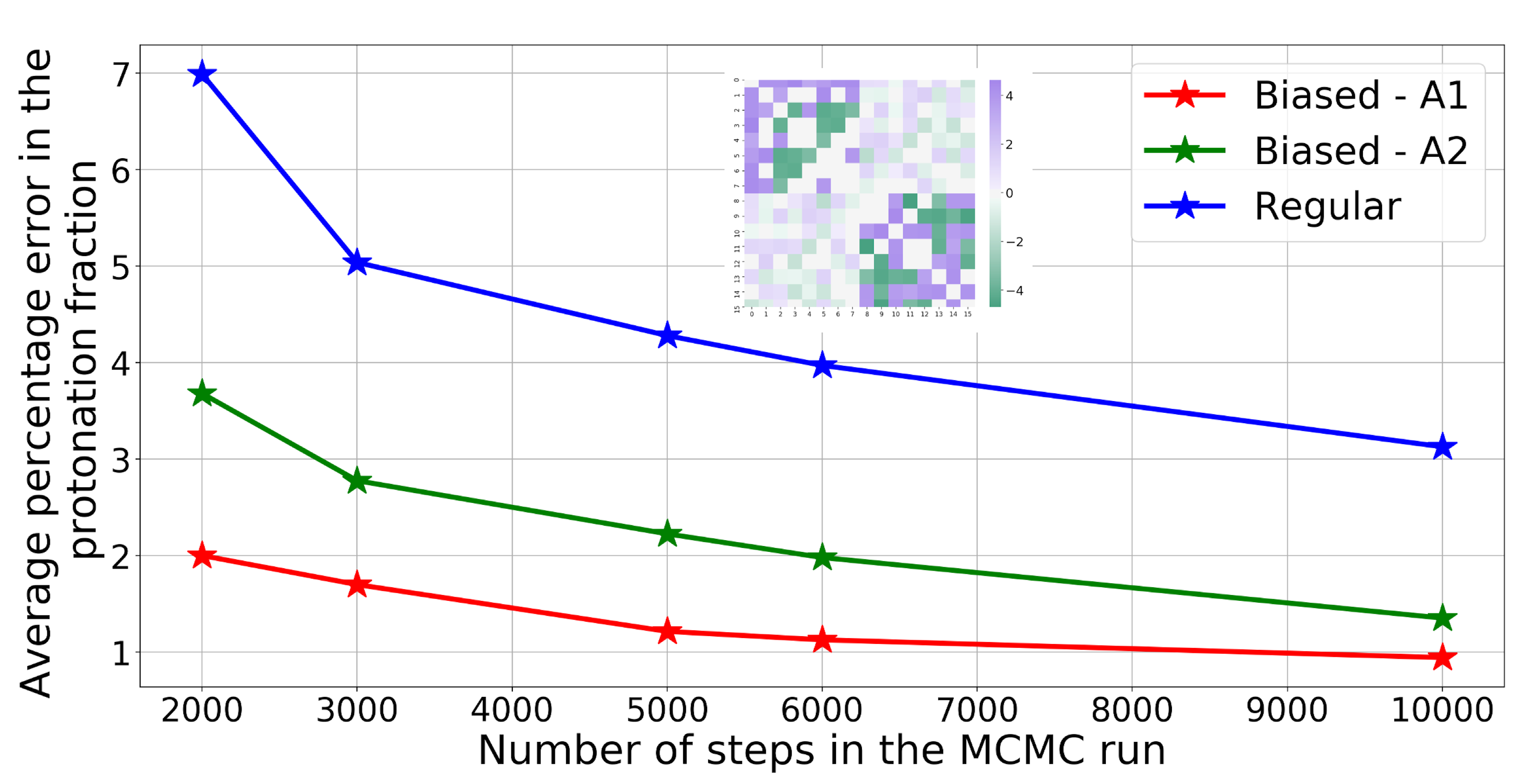}
        \caption{Error behavior comparison between regular MCMC and the proposed biased MCMC approach for System 1.
        The interaction energy matrix is shown in the inset. }
        \label{fig:synth_comp1}
    \end{centering}
\end{figure}

\begin{figure}[!htbp]
    \begin{centering}
        \includegraphics[width=0.9\columnwidth]{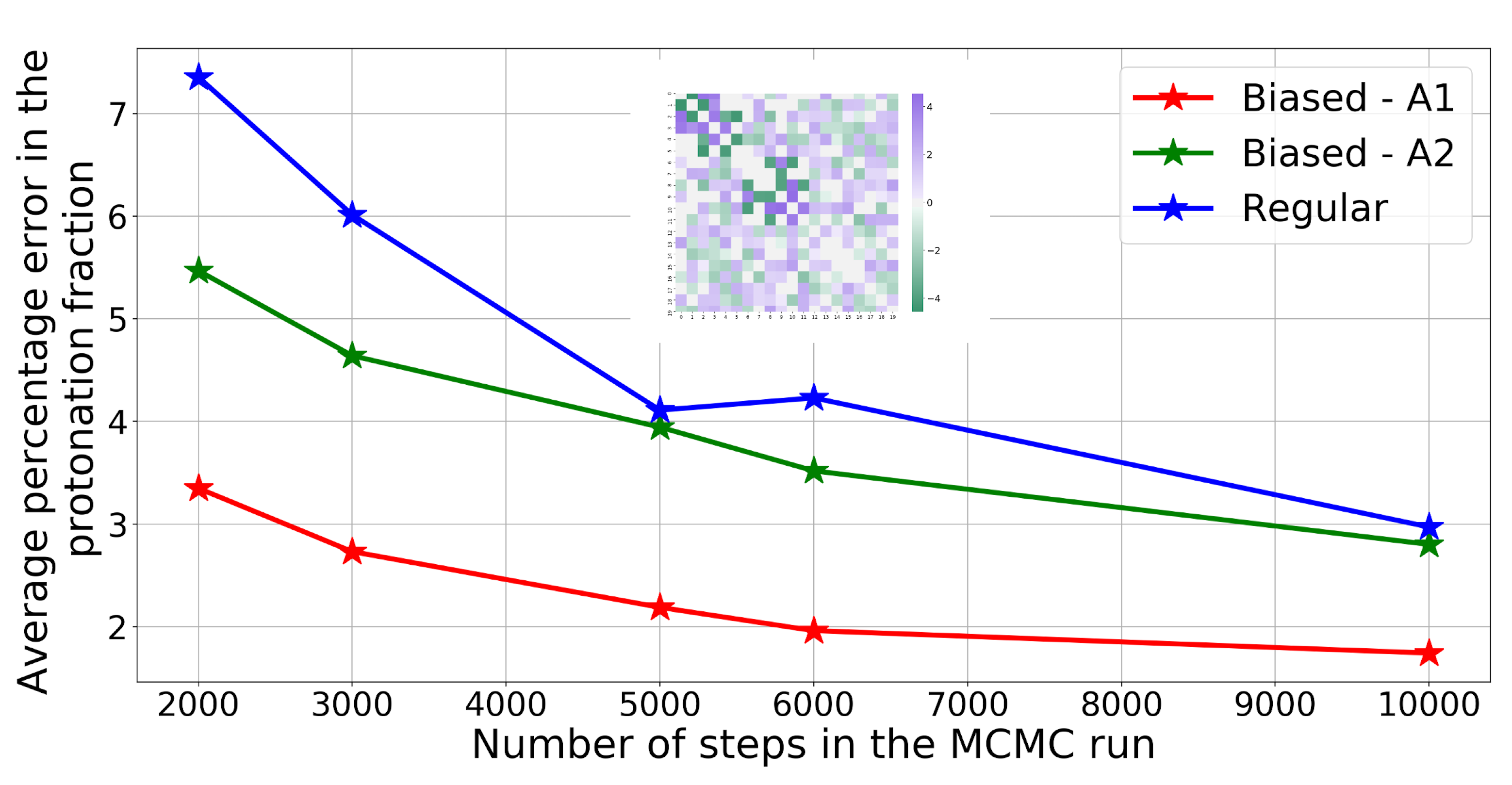}
        \caption{Error behavior comparison between regular MCMC and the proposed biased MCMC approach for System 2.
        The interaction energy matrix is shown in the inset. }
        \label{fig:synth_comp2}
    \end{centering}
\end{figure}

In order to assess the biased MCMC algorithm, we first conducted experiments on a synthesized but realistic systems titratable residues.
Synthetic systems allow us to vary system properties in a controlled manner.
The system is described by a vector of randomized self energies and a sparse matrix of interaction energies such that the interaction energy matrix can have a known density (sparsity), predefined strong and weak cluster structure, and background noise.
The first system with 16 residues had two well defined clusters of 8 residues each.
The second system also of 16 residues had two well-defined clusters of 6 residues each and 4 residues which did not form any meaningful cluster.
The visualizations of the interaction energy matrices are shown in the insets of Figures \ref{fig:synth_comp1} and \ref{fig:synth_comp2}, respectively.

We then compared the average error in the predicted protonation fractions (at a single $\pH$ value) over all the titratable residues between regular MCMC and the proposed biased MCMC algorithm using both the perturbation approaches (A1 and A2 in the figure).
We set fixed values for $k_r$: 16 and 12 for the two systems, respectively.
The small sizes of these systems allowed for the exact computation of the partition function and protonation fraction via full enumeration of all possible states.
The full enumeration method is used as the ground truth to compare the accuracy of the newly developed biased MCMC method with that of the regular MCMC.
We computed the average protonation error for a varied number of MCMC steps for both the methods.
In all cases, the burn-in phase of MCMC was fixed at 20\% of the total number of steps.   Finally, averaging was performed over multiple MCMC runs (30 in our experiments) with different initialization settings. 

For both the systems and both the approaches A1 and A2, we observed that the biased MCMC method provides significant reduction in the error over the regular MCMC method as seen in Figures \ref{fig:synth_comp1} and \ref{fig:synth_comp2} with consistently better performance from approach A1.
When compared across the two systems, we observe that the performance of the biased MCMC method slightly worsened for System 2 (with 2 strong clusters and a weak cluster) as compared to System 1 (which has only 2 strong clusters).
This is potentially attributable to the presence of the weak cluster which does not allow for the selection of the most probable $k_r$ states.
Finally, the sizes of the configuration spaces explored in biased MCMC for the two systems are approximately 0.4\% and 3.5\%  of the total configuration space sizes.
Thus, clustering effectively narrowed down the regions of the configuration space that contribute most to the partition function.

Since finding an optimal clustering is an NP-hard problem \cite{mettu_thesis}, and most clustering algorithms are polynomial-time heuristics, it is often difficult to obtain quality guarantees on the cluster assignments.
Thus, to assess the impact of clustering, we perturbed the well-defined cluster structure of System 1 by interchanging the cluster memberships of $1$,$2$, and $4$ residues at random between the two strong clusters engineered.
Table \ref{tab:clus_effect} summarizes how the accuracy of the biased MCMC method decreased as the strong clustering structure was progressively diminished.
We chose $5000$ steps for both the MCMC methods in this experiment. 
\begin{table}[!h]
    \centering
    \begin{tabular}{|| c || c | c | c | c ||} 
        \hline
        MCMC Type & $n_P=0$ & $n_P=1$ & $n_P=2$ & $n_P=4$ \\
        \hline
        Biased & 1.3\% & 2.9\% & 7.0\% & 12.5\%\\
        \hline
        Regular & 5.4\% & 5.4\%  & 5.4\% & 5.4\%  \\
        \hline
        \hline
        \end{tabular}
        \caption{The effect of sub-optimal cluster structure on the average protonation fraction error with the biased MCMC approach.}
        \label{tab:clus_effect}
\end{table}

\section{Experiments on real-world protein systems}
\label{sec:real}
We assessed the efficacy of the proposed biased MCMC methodology via experiments on two real-world protein systems.
The two proteins that we investigated were (\emph{2GUS} and \emph{4IL7}) with 14 and 12 titratable residues, respectively.
As before, these were small enough to be validated against the ground-truth answers from a complete enumeration routine.
Finally, for the real-world systems, instead of the protonation fractions, we computed the entire titration curves over a range of $\pH$ values (from $0.0$ to $20.0$ in steps of $0.5$)~\cite{Purvine2016}.
We first defined an appropriate error metric for the comparison.
Let $p$ denote the $pH$ value and let $f_r(p)$, $f_b(p)$, and $f_t(p)$ denote the curves that represent the protonation fraction of a given residue as a function of the $pH$, $p$, computed via regular MCMC, biased MCMC and full enumeration methods respectively.
The metric ($\epsilon_{xt}\left(i\right)$) is the area corresponding to the absolute difference between the titration curve for a given residue $i$ computed from a given MCMC scheme($x$) and the titration curve for residue $i$ from full enumeration which serves as the ground truth($t$):
\begin{equation}
\label{equn:error} 
\epsilon_{xt}\left(i\right) = \frac{1}{n_R}\sum_{k=1}^{k=n_R}\int_{p_L}^{p_H}\left|f_{x,k}^{i}(p)-f_t(p)\right|dp \\
\end{equation}
The bounds for the integral are defined by $pH$ values $p_L$ and $p_H$ that correspond to protonation fractions of 0.05 and 0.95 for the exact curve.
The error metric was computed by using Simpson's rule for numerical integration and  averaged by the number of independent runs $n_R$ (indexed by $k$).
We also defined an average metric $\bar{\epsilon}_{xt}$ over the number of titratable residues $N$ to facilitate comparison between methods using a single metric: $\frac{1}{N}\sum_{i=1}^{i=N}\left(\epsilon_{xt}\left(i\right) \right)$.

\begin{figure}[!htbp]
    \begin{minipage}{\columnwidth}
        \includegraphics[width=\columnwidth]{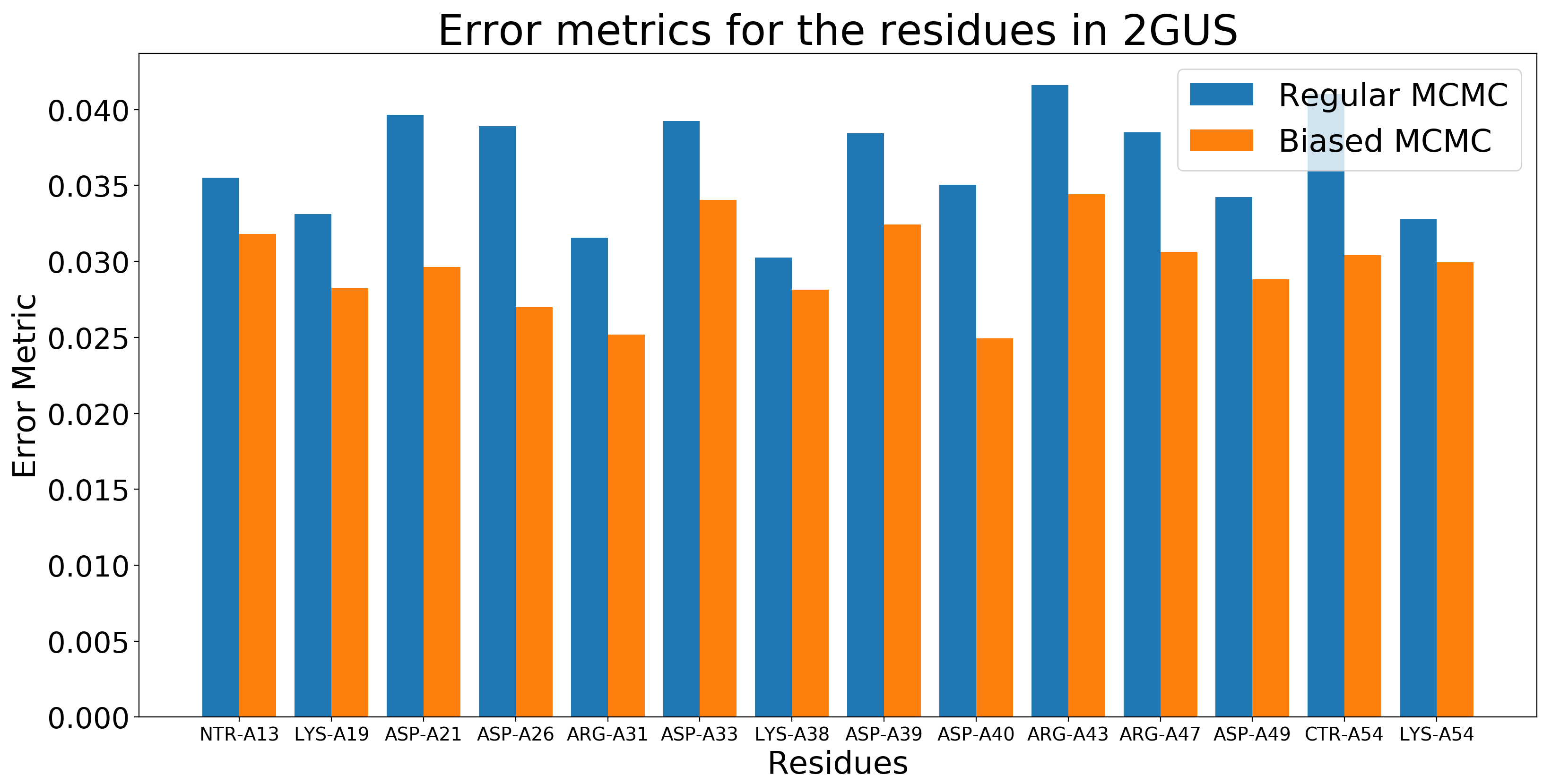}
        \caption{Error performance of the biased MCMC vs that of the regular MCMC for various titratable residues that make up the protein 2GUS. The MCMC runs used 10,000 steps and the error metrics were  averaged over 100 runs.}
        \label{fig:2gus}
    \end{minipage}
    \qquad
    \begin{minipage}{\columnwidth}
        \includegraphics[width=\columnwidth]{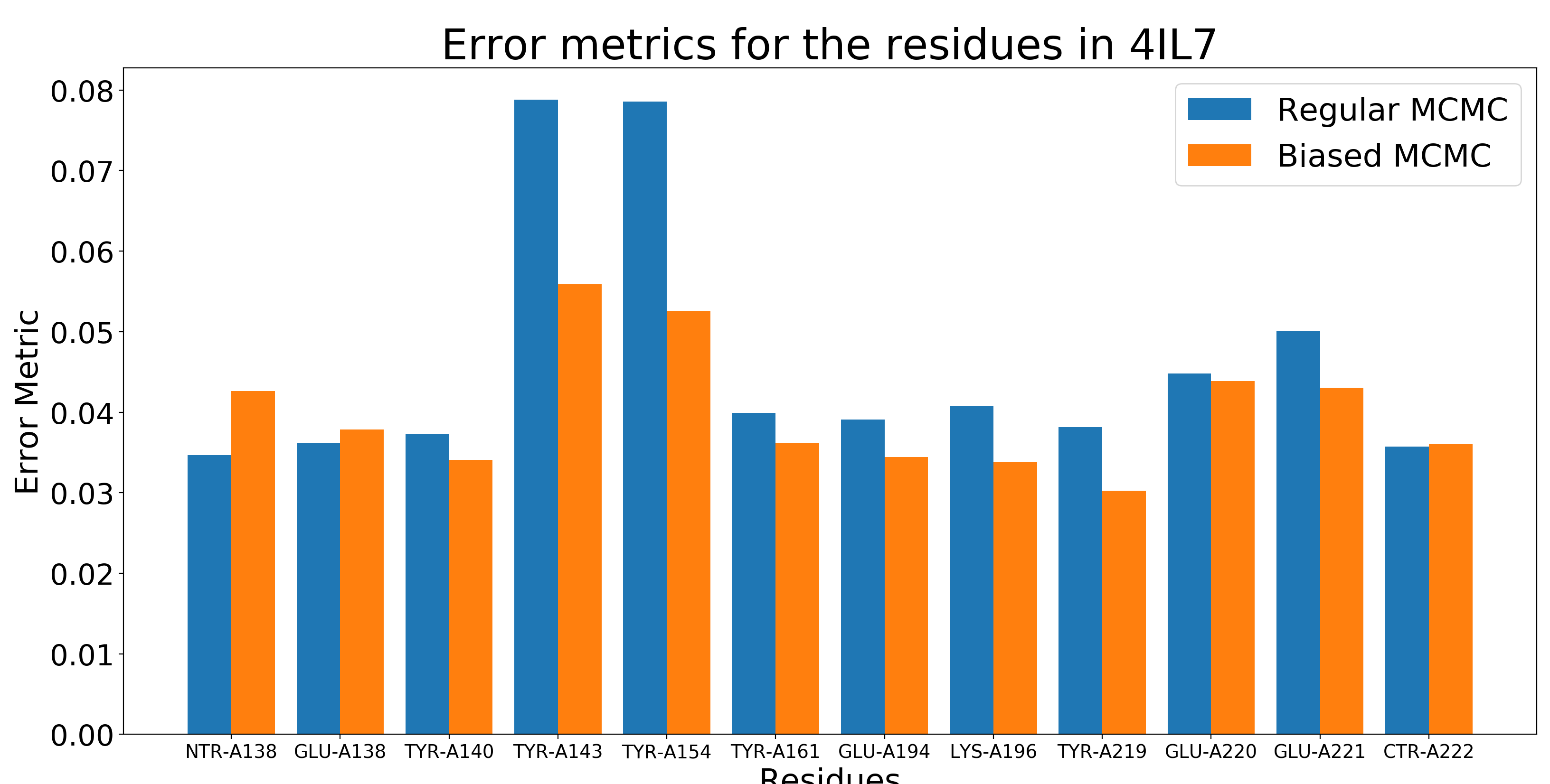}
        \caption{Error performance of the biased MCMC vs that of the regular MCMC for various titratable residues that make up the protein 4IL7. The MCMC runs used 10,000 steps and the error metrics were  averaged over 100 runs.}
        \label{fig:4il7}
    \end{minipage}
\end{figure}

Figures \ref{fig:2gus} and \ref{fig:4il7} show the comparison of the error metrics $\epsilon_{rt}$ and $\epsilon_{bt}$ for all the residues of the proteins \emph{2GUS} and \emph{4IL7}, respectively,  for each of the titratable residues.
In each of $n_R = 100$ runs, both the regular MCMC and the biased MCMC approaches used $10,000$ steps, of which, the first $2,000$ steps corresponded to the burn-in phase.
The biased MCMC method uses Approach 1 for the transitions and used 2 clusters determined by METIS and with the $k_r$ values determined by adaptive strategy described earlier.
%

Figures \ref{fig:2gus} and \ref{fig:4il7} also show that the biased MCMC approach was more accurate for nearly all the residues.
Furthermore, a small number of residues showed significant improvement in accuracy over the regular MCMC method.
Table \ref{tab:avg_error_residue} shows the percentage improvement in the error metric $\bar{\epsilon}_{bt}$, for the biased MCMC over the error metric for regular MCMC, namely,  $\bar{\epsilon}_{rt}$.
$R$ is the interval for protonation fractions from which the limits of the integral (in terms of $pH$ values) were computed.
The relative performance gain of the biased MCMC method was slightly better for the larger interval. 

While these results showed a noticeable improvement in the error performance of the biased MCMC method over the regular MCMC method, the improvement was less significant when compared to the synthetic systems with stronger cluster structure.
This investigation is part of our ongoing work.
\begin{table}[!h]
    \begin{center}
        \begin{tabular}{|| c || c | c || c | c ||} 
            \hline
            Protein & $R$=[0.05,0.95], & $R$=[0.01,0.99] \\
            \hline
            2GUS & -18.7\% & -19.5\% \\
            \hline
            4IL7 & -13.4\% & -15.9\% \\
            \hline
            \hline
        \end{tabular}
        \caption{Relative improvement in the error metrics for biased MCMC over regular MCMC for two real-world proteins.}
        \label{tab:avg_error_residue}
    \end{center}
\end{table}

\section{Conclusions}
\label{sec:conclu}
In this work, we presented a novel biased MCMC algorithm for computing ensemble averages in systems characterized by pair-wise interactions and specifically protein residue networks.
The algorithm exploited the cluster structure of the interaction energy matrix and allowed the residue protonation states to change together.
We showed that our new schemes represent valid MCMC runs that satisfy ergodicity and detailed balance requirements.
We implemented our biased MCMC algorithms and applied them to a number small protein residue network systems, both synthetic and real-world and showed that our new schemes provide improved accuracy in the protonation fraction and hence titration curve estimates.
We further showed the sensitivity of the methods to the quality of the clustering assignments.
Our future work will focus on better clustering strategies to further improve the performance on real-world systems, optimizing the hyper-parameters and application of the method to larger real-world protein residue networks. 

\section{Acknowledgments}
This work was funded by NIH grant R01 GM069702.

\balance
\bibliographystyle{IEEEtran}
\bibliography{biasedmc}

\end{document}